\documentclass[aps,pre,citesort]{revtex4}
\usepackage[dvips]{graphicx}
\usepackage{latexsym}

\usepackage{fancyheadings}
\begin{document}

\title{The Effect of Nanoparticle Shape on \\ Polymer-Nanocomposite Rheology and Tensile Strength}

\author{Scott T. Knauert$^{1}$, Jack F. Douglas$^{2,}$\footnote{Official Contribution of NIST - Not Subject to Copyright in the United States}, and Francis W. Starr$^{1}$} 
\affiliation{$^{1}$Department of Physics, Wesleyan University, Middletown CT 06459, USA}
\affiliation{$^{2}$Polymers Division, National Institute of Standards and Technology, Gaithersburg, Maryland 20899, USA}
\begin{abstract}

Nanoparticles can influence the properties of polymer materials by a variety of mechanisms. With fullerene, carbon nanotube, and clay or graphene sheet nanocomposites in mind, we investigate how particle shape influences the melt shear viscosity $\eta$ and the tensile strength $\tau$, which we determine via molecular dynamics simulations. Our simulations of compact (icosahedral), tube or rod-like, and sheet-like model nanoparticles, all at a volume fraction $\approx 0.05$, indicate an order of magnitude increase in the viscosity $\eta$ relative to the pure melt. This finding evidently can not be explained by continuum hydrodynamics and we provide evidence that the $\eta$ increase in our model nanocomposites has its origin in chain bridging between the nanoparticles. We find that this increase is the largest for the rod-like nanoparticles and least for the sheet-like nanoparticles. Curiously, the enhancements of $\eta$ and $\tau$ exhibit {\it opposite trends} with increasing  chain length $N$ and with particle shape anisotropy. Evidently, the concept of bridging chains alone cannot account for the increase in $\tau$ and we suggest that the deformability or flexibility of the sheet nanoparticles contributes to nanocomposite strength and toughness by reducing the relative value of the Poisson ratio of the composite. We note the molecular dynamics simulations in the present work focus on the reference case where the modification of the melt structure associated with glass-formation and entanglement interactions should not be an issue. Since many applications require good particle dispersion, we also focus on the case where the polymer-particle interactions favor nanoparticle dispersion. Our simulations point to a substantial contribution of nanoparticle shape to both mechanical and processing properties of polymer nanocomposites.

\end{abstract}

\date{\today}

\maketitle

\section{Introduction}

Blends of polymers and nanoparticles, commonly called ``polymer nanocomposites'' (PNC), have garnered much attention due to the possibility of dramatic improvement of polymeric properties with the addition of a relatively small fraction of nanoparticles~\cite{Wypych,Giannelis,Roco_Williams_Alivisatos,Crandall,Vaia_Giannelis,Vaia_Giannelis_Book,Starr_Glotzer}. Successfully making use of these materials depends upon a firm understanding of both their mechanical and flow properties. Numerous computational and theoretical studies have examined the clustering and network formation of nanoparticles and their effect on both the structural and rheological properties of PNCs~\cite{Starr_Douglas_Glotzer,Starr_Schroder_Glotzer_2001,Starr_Schroder_Glotzer_2002,Kobelev_Schweizer,Chen_Kobelev_Schweizer,Hooper_Schweizer_2005,Hooper_Schweizer_2006,Hooper,Ganesan_Pryamitsyn_Surve_Narayanan,Surve_Pryamitsyn_Ganesan,Ganesan_Macromolecules,Ganesan_Journal_of_Rheology,Vacatello,Vacatello_Macromolecules,Salaniwal_Kumar_Douglas,Sen_Kumar_Keblinski,Desai_Keblinski_Kumar,Sinsawat_Anderson_Vaia_Farmer,Kairn}. The vast majority of these efforts have focused on nanoparticles that are either spherical, polyhedral or otherwise relatively symmetric, although there are some notable exceptions~\cite{Surve_Pryamitsyn_Ganesan,Ganesan_Macromolecules,Ganesan_Journal_of_Rheology,Sinsawat_Anderson_Vaia_Farmer}. In contrast, experiments have tended to emphasize highly asymmetric nanoparticles~\cite{Lebaron_Wang_Pinnavaia,Lan_Pinnavaia,Ray_Okamoto,Kumar,Potschke_Fornes_Paul,Mamedov,Goel,Ren_Krishnamoorti,Hsied,Xu,Osman_2005,Osman_2006}, such as layered silicates or carbon nanotubes. It is generally appreciated that these highly asymmetric nanoparticles have the potential to be even more effective than spherical (or nearly spherical) nanoparticles in changing the properties of the polymer matrix to which they are added. In addition to the large enhancements in viscosity and shear modulus expected from continuum hydrodynamic and elasticity theories, extended nanoparticles can more easily form network structures both through direct interaction between the nanoparticles, or through chain bridging between the nanoparticles~\cite{Hooper_Schweizer_2005,Hooper_Schweizer_2006,Salaniwal_Kumar_Douglas,Gersappe}, where a ``bridging'' chain is a chain in contact with at least two different nanoparticles. These non-continuum mechanisms are believed to play a significant role in property enhancement, though the dominant mechanism depends on the properties considered, particle-polymer and particle-particle interactions, sample preparation, etc. 

Given that the majority of previous computational efforts have focused on symmetric nanoparticles, we wish to elucidate the role of {\it nanoparticle shape} in determining basic material properties, such as the viscosity $\eta$, and material ``strength'', (i.e., breaking stress). Computer simulations are well suited to examine the role of nanoparticle shape, since it is possible to probe the effects of changing the shape without the alteration of any of the intermolecular interactions. In this way, the changes due to nanoparticle shape can be isolated from other effects. Such a task is complicated experimentally, since it is difficult to modify the shape of a nanoparticle without dramatically altering its intermolecular interactions. In this paper we evaluate the viscosity $\eta$ and ultimate isotropic tensile strength $\tau$ of model PNC systems with either (i) highly symmetric icosahedral nanoparticles (compact particles), (ii) elongated rod-like nanoparticles, and (iii) sheet-like nanoparticles. These nanoparticles can be thought of as idealizations of common nanoparticles, such as gold nanoparticles and fullerenes (polyhedral), nanotubes and fibers, and nanoclay and graphene sheet materials, respectively. Our results are based on molecular dynamics (MD) computer simulations, using non-equilibrium methods to evaluate $\eta$~\cite{Frenkel_Smit,Allen_Tildesley}, and exploiting the ``inherent structure'' formalism to determine $\tau$~\cite{Sastry_Debendetti_Stillinger,Shen_Debenedetti_Stillinger}. We find that the rod-like nanoparticles give the largest enhancement to $\eta$, which we correlate with the presence of chains that bridge between the nanoparticles. The sheet nanoparticles offer the weakest increase in $\eta$, and correspondingly have the smallest fraction of bridging chains. For the ultimate isotropic strength $\tau$, we find opposite results: the sheets provide the greatest reinforcement, while the rods the least. For both of these properties, the property changes induced by the icosahedral nanoparticles fall between those of the extended nanoparticles.

The present simulations are idealized mixtures of polymers and nanoparticles in which the polymer-nanoparticle interactions are highly favorable so as to promote nanoparticle dispersion. Moreover, we have chosen to work at relatively high temperature in order to avoid contributions to $\eta$ from the complex physics of slowing dynamics approaching the glass transition. Previous work~\cite{Starr_Schroder_Glotzer_2001,Starr_Schroder_Glotzer_2002} has shown that polymer-surface interaction effects in this low temperature range can alter, and potentially dominate the nanocomposite properties. We also limit the range of chain length $N$ studied to avoid effects of significant polymer entanglement. These limitations on interaction, temperature, and chain length are advantageous in order to develop a clear understanding of the origin of the observed changes in properties. Such a reference calculation provides a reference starting point to understand behavior when these constraints are relaxed. With this in mind, caution is needed when comparing these results with experimental data where these complicating additional factors may be present -- along with other possible effects, such as crystallization or phase separation. 

We organize this paper as follows: in Section~\ref{sec:Simulation}, we describe the details of the model and method, focusing on the differences between the nanoparticle types used in each system. Section~\ref{sec:Composite_Rheology} describes our investigation of the rheological properties of the nanocomposites, while Section~\ref{sec:Isotropic_Tensile_Strength} considers the effects of shape on $\tau$. We conclude in Section~\ref{sec:Conclusion}. 

\section{Simulation}
\label{sec:Simulation}

To directly compare to experiments, it is desirable to use as realistic a molecular model as possible. While a chemically accurate MD simulation is possible in principle, it is often more difficult to identify basic physical trends with such models. Such attempts at chemical realism are also demanding in terms of the computational times required, which restricts the class of problems which can be investigated. Coarse-grained models of polymeric materials provide a good compromise between the the opposing needs of realism and computational feasibility.  Such models can reproduce qualitative experimental trends of nanocomposites, but precise quantitative predictions cannot be expected~\cite{Kairn}. Building on nanocomposite models introduced before~\cite{Starr_Douglas_Glotzer,Starr_Schroder_Glotzer_2001,Starr_Schroder_Glotzer_2002,Smith_Bedrov_Li_Byutner}, we study a coarse-grained model for polymers~\cite{Rudisill_Cummings,Grest_Kremer} and nanoparticles that allow us to consider PNC systems over a wide range of physically interesting systems. Here we consider several nanoparticle shapes built by connecting spherical force sites.

We perform MD simulations of systems consisting of a small fraction of model nanoparticles in a dense polymer melt. For reference we also simulate the corresponding pure polymer melt. The polymers are modeled via the common ``bead-spring'' approach, where polymers are represented by chains of monomers (beads) connected by bond potentials (springs)~\cite{Doi_Edwards}. All monomers interact via a modified Lennard Jones (LJ) potential
\begin{equation}
V_{\rm P-P}(r_{\rm ij}) = \left\{ \begin{array} {r@{\quad:\quad}l} 4 \epsilon_{\rm P-P}\left[ \left(\frac{\displaystyle \sigma}{\displaystyle r_{\rm ij}}\right)^{12} - \left(\frac{\displaystyle \sigma}{\displaystyle r_{\rm ij}}\right)^6 \right] - V_{\rm LJ}(r_c) & r_{\rm ij} \le r_c \\ 0 & r_{\rm ij} > r_c \end{array} \right.,
\label{Equation:VLJ}
\end{equation}
where $r_{\rm ij}$ is the distance between two monomers, where $\epsilon_{\rm P-P}$ is the depth of the well of the LJ potential and $\sigma$ is the monomer size. The potential is truncated and shifted at $r_c = 2.5~\sigma$, so that the potential and force are continuous at the cutoff. Bonded monomers in a chain interact via a finitely extensible, non-linear elastic (FENE) spring potential~\cite{Rudisill_Cummings,Grest_Kremer}
\begin{equation}
V_{\rm FENE}(r_{\rm ij}) = -k\left(\frac{R_0^2}{2}\right) \ln \left[1-\left(\frac{r_{\rm ij}}{R_0}\right)^2\right],
\label{Equation:VFENE}
\end{equation}
where $k$ and $R_0$ are adjustable parameters that have been chosen as in ref.~\cite{Rudisill_Cummings,Grest_Kremer}. Since we do not aim to study a specific polymer, we use reduced units where $\sigma = \epsilon_{\rm P-P} = m = 1$ ($m$ is the monomer mass). Length is defined in dimensionless units relative to $\sigma$, time in units of $\sigma \sqrt{m/\varepsilon}$ and temperature $T$ is expressed in units of $\varepsilon/k_B$, where $k_B$ is Boltzmann's constant. 

\begin{figure}[t]
\begin{center}
\includegraphics[clip,width=5.5cm]{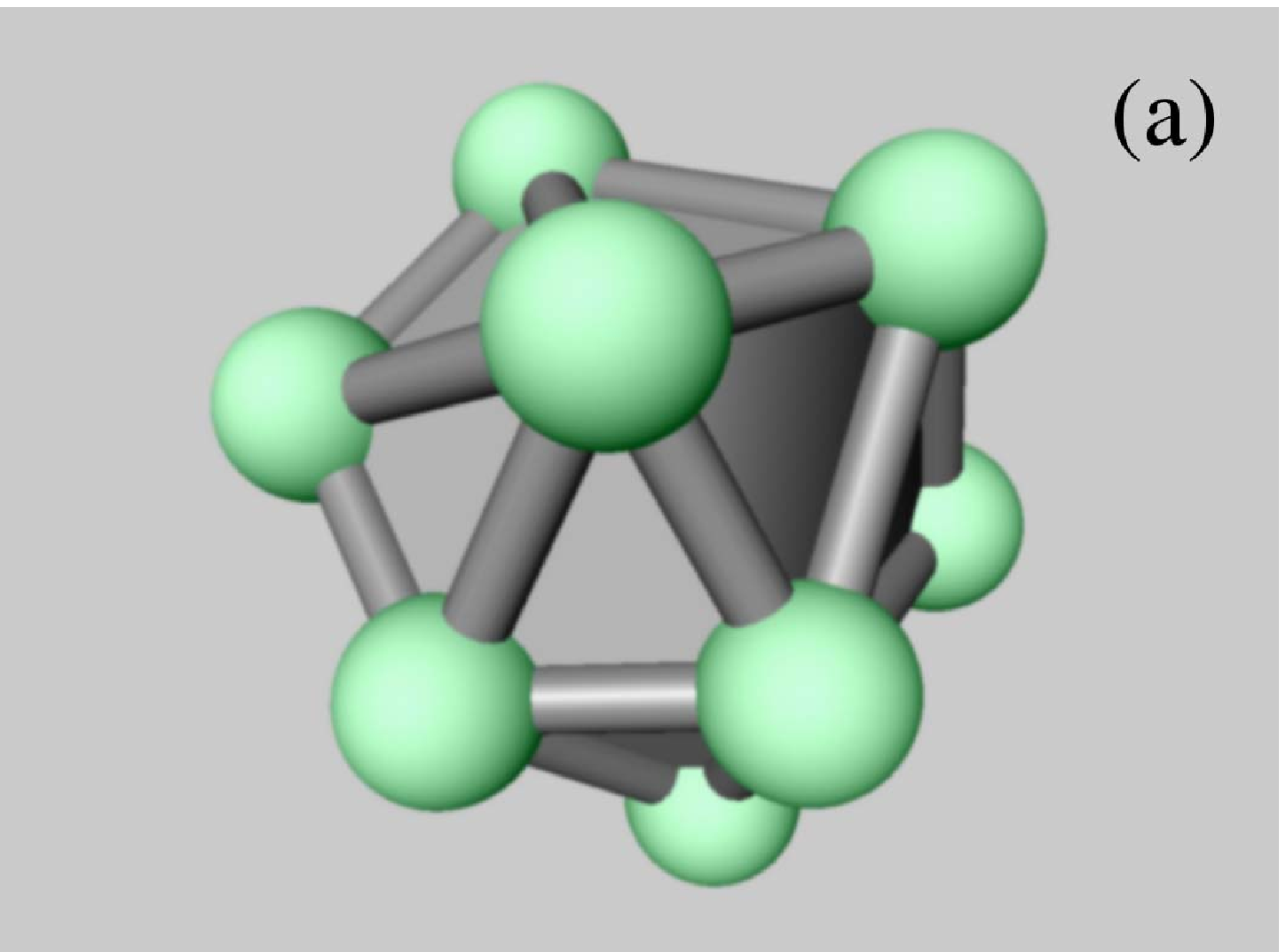}
\includegraphics[clip,width=5.5cm]{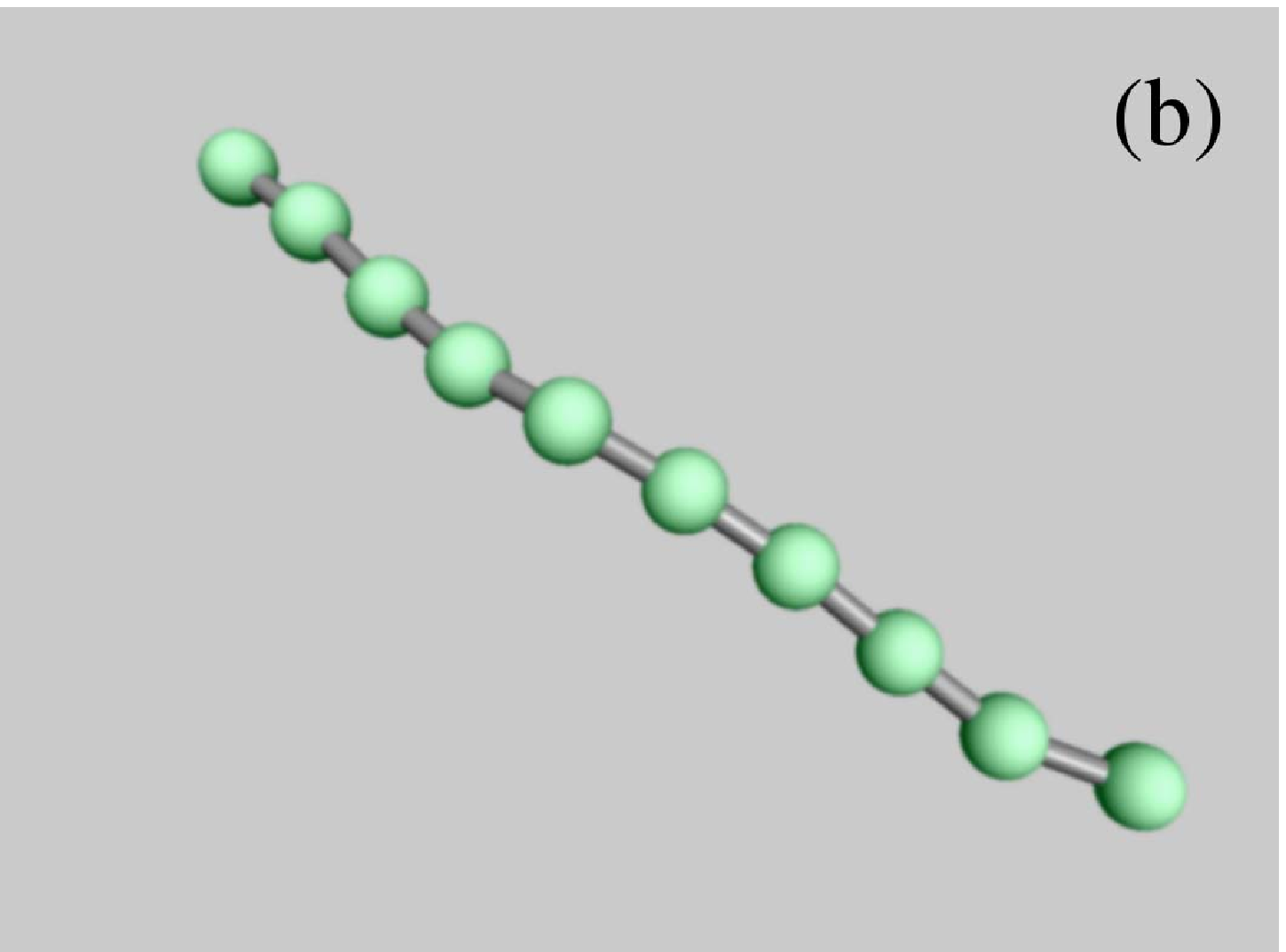}
\includegraphics[clip,width=5.5cm]{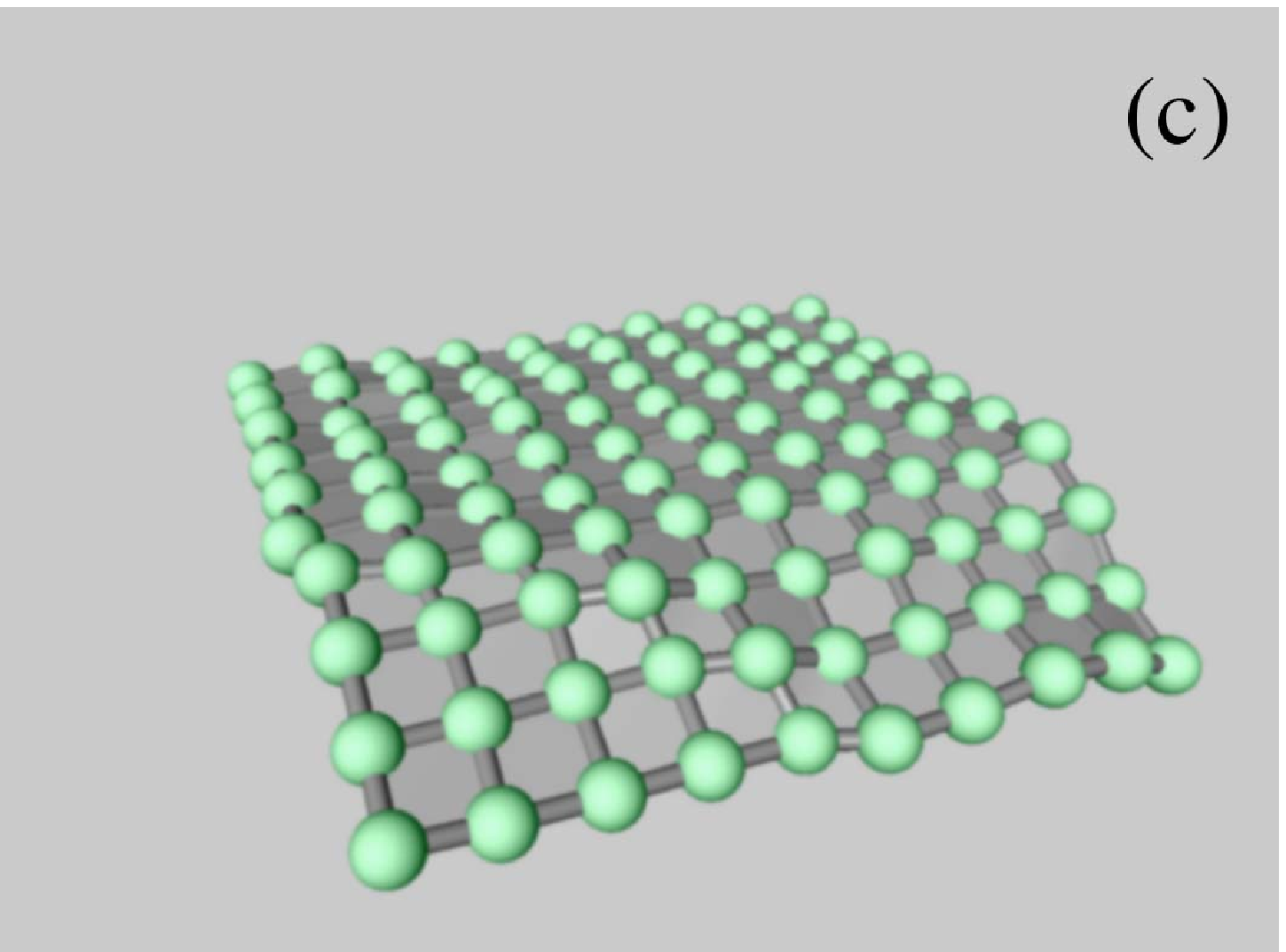}
\end{center}
\caption{The three different types of nanoparticles used in simulation: (a) icosahedron, (b) rod, and (c) sheet. The nanoparticle force sites are rendered as spheres connected by cylinders representing FENE bonds.}
\label{fig:Nanoparticles} 
\end{figure}

We use three different types of nanoparticles for our calculations. Fig.~\ref{fig:Nanoparticles} shows representative images of the nanoparticles. The first type of nanoparticle, an icosahedron, was previously studied in ref.~\cite{Starr_Douglas_Glotzer,Starr_Schroder_Glotzer_2001,Starr_Schroder_Glotzer_2002} which focused on factors controlling nanoparticle dispersion and low temperature effects on transport for a similar polymer matrix, respectively. Practical realizations include fullerene particles, primary carbon black particles, quantum dots, and metal nanoparticle additives. The nanoparticle force sites interact with each other via an identical $V_{\rm P-P}$ given in Eq.~(\ref{Equation:VLJ}), with $\sigma = \epsilon_{\rm P-P} = m = 1$. To maintain the icosahedral shape, the force site at each vertex is bonded to its 5 nearest neighbors via a harmonic spring potential
\begin{equation}
V_{\rm harm}(r_{\rm ij}) = -\kappa \frac{r_0^2}{2}\left(\frac{r_{\rm ij}}{r_0}-1\right)^2
\label{Equation:VHARM}
\end{equation}
where $\kappa = 60$ and $r_0$ equals the minimum of the force-shifted LJ potential, approximately $2^{1/6}$. To further reinforce the icosahedral geometry, a central particle is bonded to the vertices with the same potential Eq.~(\ref{Equation:VHARM}) with the same value of $\kappa$, but with a slightly smaller preferred bond length, $r_0' = 1/4 (10+2\sqrt{5})^{1/2}r_0$, the radius of the sphere circumscribed around an icosahedron. The resulting nanoparticles have some flexibility, but are largely rigid, thereby preserving their icosahedral shape.

The second type of nanoparticle is a semiflexible rod represented by 10 LJ force sites with neighboring monomers bonded by the same $V_{\rm FENE}$ as used for the polymers. We choose the shape to represent nanoparticles such as carbon nanotubes or nanofibers. Carbon nanotubes typically have lengths up to several $\mu$m and a diameter of 1~nm to 2~nm (single-walled tubes) or 2~nm to 25~nm (multiwalled tubes can have even larger diameters)~\cite{Ajayan_Ebbesen}. Unfortunately, such large rods are not feasible to simulate, as the system size needed to avoid finite size effects exceeds current computational resources. As a compromise, we simulate rods with a length-to-diameter ratio of 10. There is an additional bond potential between nanoparticle force sites 
\begin{equation}
V_{\rm lin}(\theta) = \kappa_{\rm lin}(1+\cos{\theta})
\label{Equation:VLIN}
\end{equation}
where $\theta$ is the angle between three consecutive force sites. This imparts stiffness to the rod. We choose $\kappa_{\rm lin}=50$, so that like the icosahedra, the rods have some flexibility, but are largely rigid.

The third type of nanoparticle is square ``sheet'' comprised of 100 force sites. While the sheets are obviously different from the rods, the sheets have the same aspect ratio as the rods, where aspect ratio is defined by the ratio of the largest and smallest length scales. Aspect ratio is often considered to be one of the most important properties for anisotropic particles, apart from the interactions with the surrounding matrix. The sheets represent a coarse-grained model of clay-silicate nanoparticles~\cite{Sinsawat_Anderson_Vaia_Farmer}, and in the literature these objects are also termed ``tethered membranes''~\cite{Starr_Schroder_Glotzer_2001,Starr_Schroder_Glotzer_2002,Rudisill_Cummings,Grest_Kremer,Douglas}. Each monomer in the $10\times10$ array has non-bonded interactions described by the same LJ forces as the polymers. The monomers are also bonded to their 4 nearest neighbors via the same FENE bond potential of Eq.~(\ref{Equation:VLIN}). The particles at edges and corners of the sheet are only bonded to 3 and 2 neighboring particles, respectively. Like the rods, the sheets are stiffened with a $V_{\rm lin}$ potential to prevent the nanoparticle from folding in on itself. The sheets also include a perpendicular bonding potential 
\begin{equation}
V_{\perp}(\theta) = \frac{k_\perp}{2} \left(\theta - \frac{\pi}{2}\right)^2
\label{Equation:VPERP}
\end{equation}
which limits distortions from a square geometry. Without the potential $V_{\perp}$, the sheet can deform into a rhombus. As in ref.~\cite{Sinsawat_Anderson_Vaia_Farmer}, we choose $\kappa_{\rm lin}=10$ and $k_{\perp}=100$ for the sheets.

\begin{figure}[t]
\begin{center}
\includegraphics[clip,width=5.5cm]{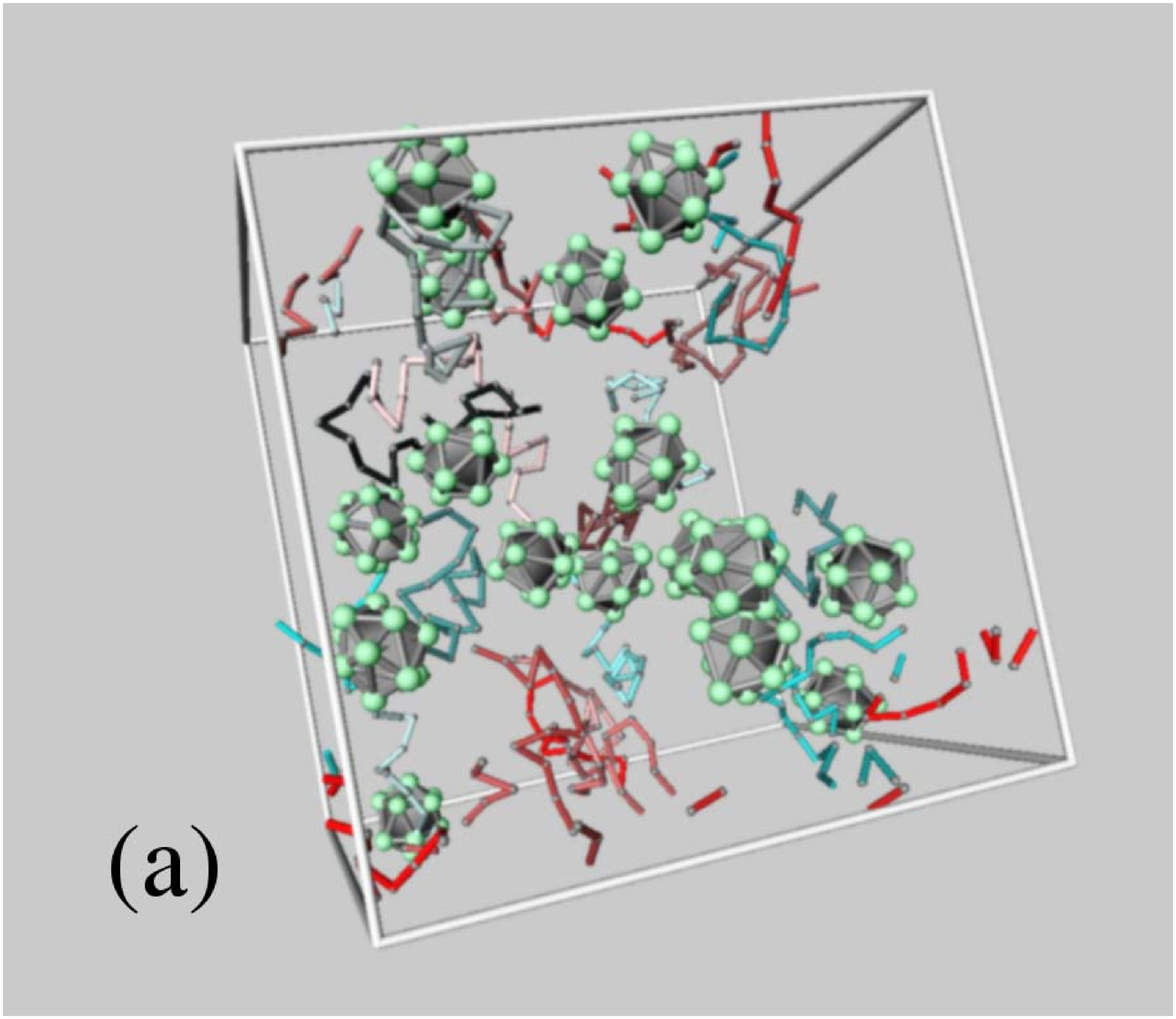}
\includegraphics[clip,width=5.5cm]{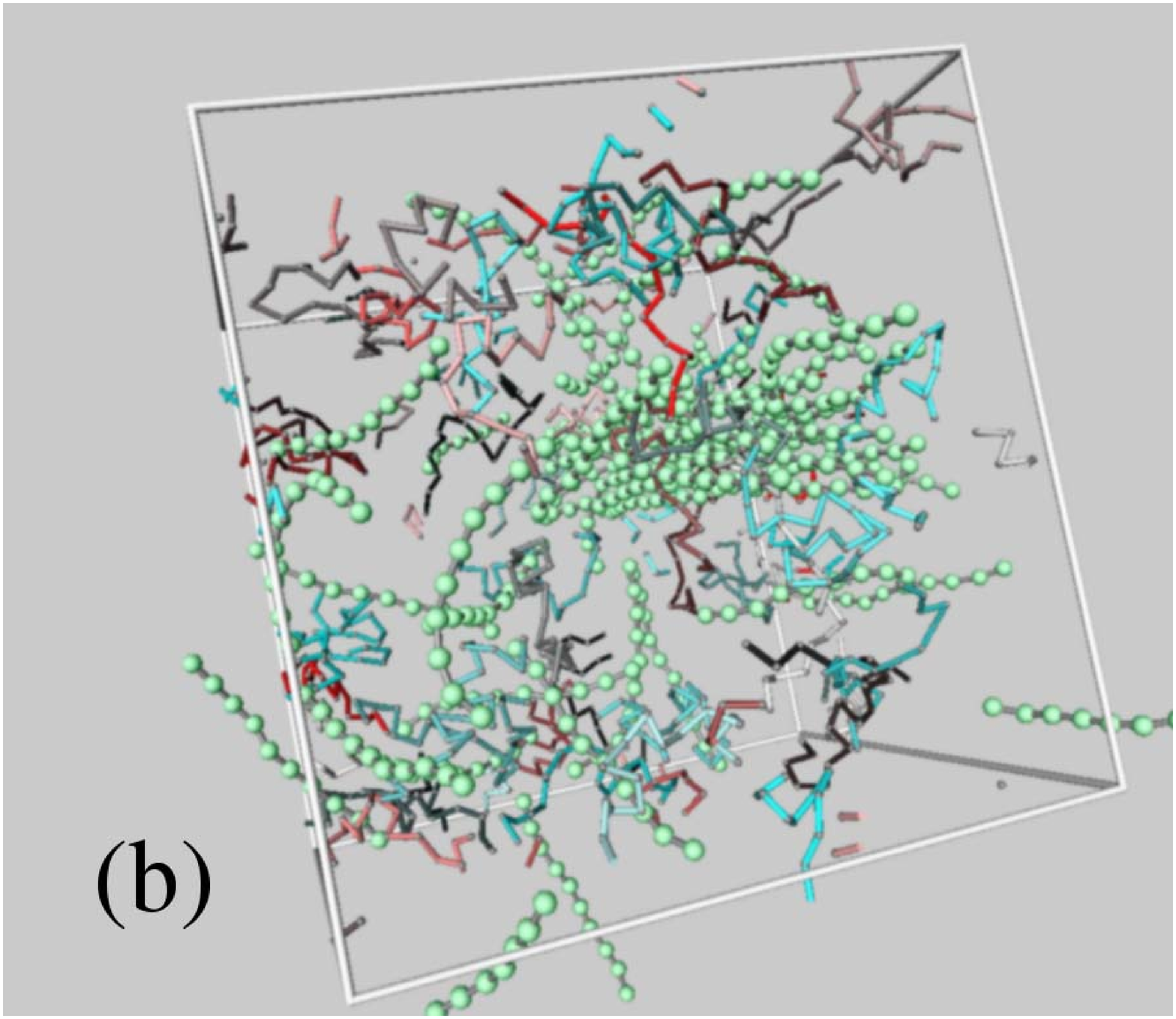}
\includegraphics[clip,width=5.5cm]{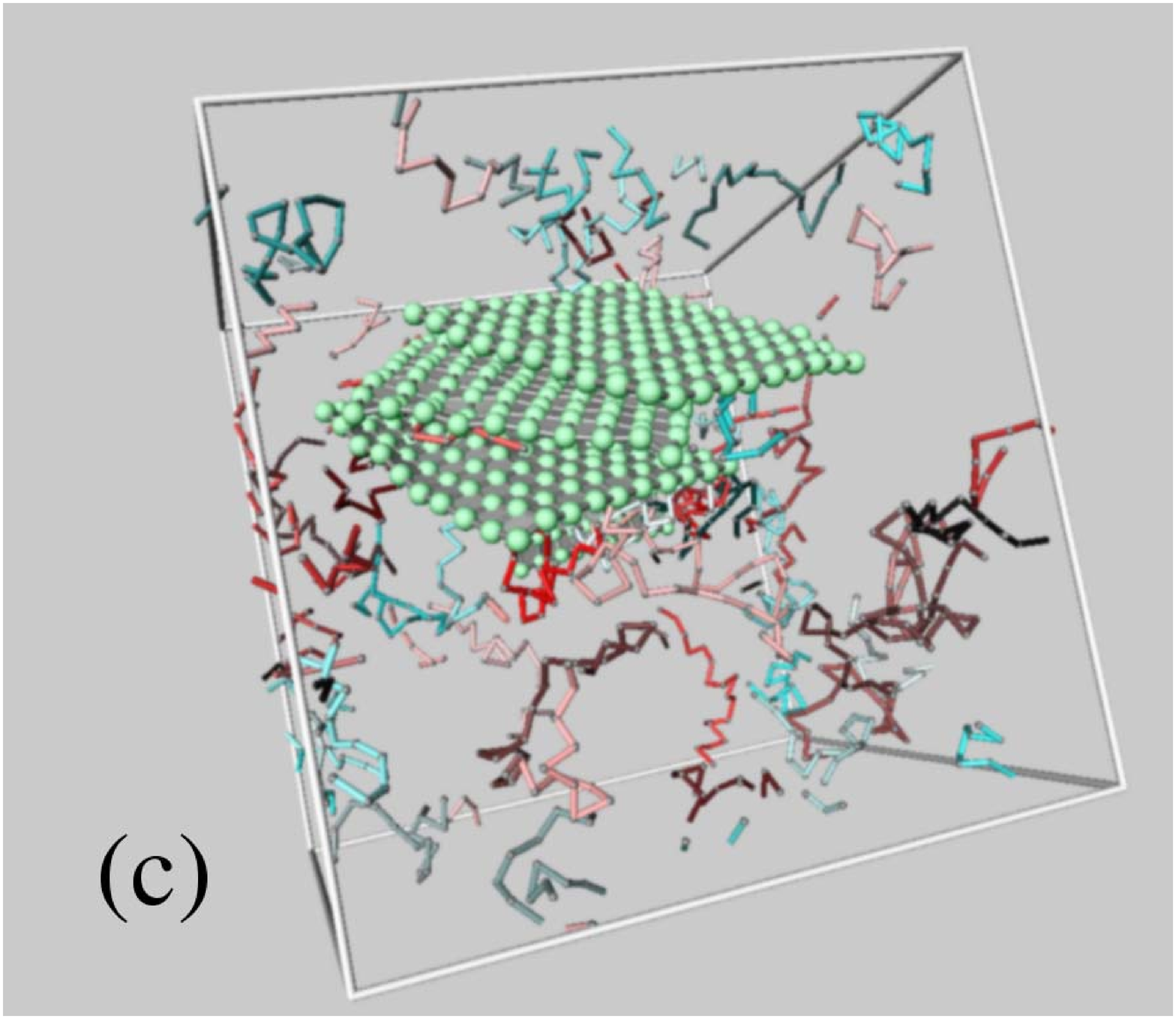}
\end{center}
\caption{Typical equilibrium configurations of each of the three nanocomposite systems: (a) icosahedra, (b) rods, and (c) sheets. In each image all nanoparticles are shown; for clarity of the figure, only 10~\% of the polymers are shown. }
\label{fig:Systems}
\end{figure}

Thus far, we have not defined a nanoparticle-monomer potential. In previous work~\cite{Starr_Douglas_Glotzer,Starr_Schroder_Glotzer_2001,Starr_Schroder_Glotzer_2002}, the same LJ potential of Eq.~(\ref{Equation:VLJ}) was used for nanoparticle-monomer interactions, with the well depth $ \epsilon_{\rm P-P}$ replaced by $\epsilon_{\rm N-P}$. The soft $\frac{1}{r^6}$ form of the attraction makes it relatively easy for monomers which are first and second nearest neighbors of a nanoparticle to exchange. As a result, the chains can ``slide'' relatively easily along the nanoparticle surface, thereby reducing any benefits of networks built by chain bridging~\cite{Kumar_FN}. To make monomer exchange at the surface less favorable, we use a ``12-24'' potential $V_{12-24}$ rather than the standard ``6-12'' powers of $V_{\rm LJ}$. Thus, the the nanoparticle-monomer interactions are of the form
\begin{equation}
V_{\rm N-P}(r_{\rm ij}) = \left\{ \begin{array} {r@{\quad:\quad}l} 4 \epsilon_{\rm N-P}\left[ \left(\frac{\displaystyle \sigma}{\displaystyle r_{\rm ij}}\right)^{24} - \left(\frac{\displaystyle \sigma}{\displaystyle r_{\rm ij}}\right)^{12} \right] - V_{12-24}(r_c) & r_{\rm ij} \le r_c \\ 0 & r_{\rm ij} > r_c \end{array} \right.,
\label{Equation:DeepWell}
\end{equation}
where $\sigma = 1$ and $r_c = 2.5\; \sigma$. This potential has shorter range and stronger forces binding the first neighbor monomers to the nanoparticle. So that the total energy of the potential well, i.e. $4\pi\int^{r_c}_\sigma V_{\rm N-P} \mathrm{d}r$, is the same as that of the LJ potential used in ref.~\cite{Starr_Schroder_Glotzer_2001,Starr_Schroder_Glotzer_2002}, we choose $\epsilon_{\rm N-P} = 3$. In this way, the total potential energy is comparable to that used in ref.~\cite{Starr_Schroder_Glotzer_2001,Starr_Schroder_Glotzer_2002} with $\epsilon_{\rm N-P}=1.5$, which led to well-dispersed nanoparticles in that study. Thus, we expect our systems will be well-dispersed. However, we shall see this is {\it not} entirely the case for the systems with sheet nanoparticles. Figure~\ref{fig:Systems} shows snapshots of a typical configuration for each system studied.

We generate initial configurations using the same approach as in ref.~\cite{Starr_Schroder_Glotzer_2001,Starr_Schroder_Glotzer_2002}, namely growing vacancies so that the nanoparticles can be accommodated. However, this process generates artificial initial configurations, and so we generate subsequent ``seed'' configurations by simulating at $T=5$ where reorganization occurs on relatively short time scales. Depending on the nanoparticle type, we extract independent starting configurations every $10^5$ to $10^6$ time steps. Subsequently, we make the necessary changes in density and chain length before cooling and relaxing at $T=2$. A possible concern is that at $T=2$, monomers will stick to the nanoparticle surface on a time scale that is long compared to the simulation, since $\epsilon_{\rm N-P} = 3$; however, we confirmed that monomers of chains exchange with the nanoparticle surface many times over the simulation. The equilibration time depends on the system type and is directly related to particle size. Pure systems relax relatively quickly, needing only $10^5$ time steps. Icosahedral systems need roughly $10^6$ time steps before reaching thermodynamic equilibrium. The rods and sheet nanoparticles are much larger, and thus diffuse more slowly. As a result, these systems require in excess of $5\rm \times 10^6$ time steps to reach equilibrium. Given that equilibration requires more than $10^6$ time steps at $T=2$, equilibration under conditions of entanglement or supercooling would be computationally prohibitive.

We integrate the equations of motion via the reversible reference system propagator algorithm (rRESPA), a multiple time step algorithm used to improve simulation speed~\cite{Frenkel_Smit}. We use a basic time step of 0.002 for a 3-cycle velocity Verlet version of rRESPA with forces divided into ``fast'' bonded ($V_{\rm FENE}$, $V_{\rm harm}$, $V_{\rm lin}$, $V_{\perp}$) and ``slow'' non-bonded ($V_{\rm P-P}$,$V_{\rm N-P}$) components. The temperature is controlled using the Nose-Hoover method where the adjustable ``mass'' of the thermostat is selected to match the intrinsic frequency obtained from theoretical calculations of a face-centered cubic LJ system~\cite{Cui_Cummings_Cochran}.

To study rheological properties, we shear equilibrated configurations using the SSLOD equations of motion~\cite{Cui_Cummings_Cochran}, integrated using the same rRESPA algorithm used for equilibrium simulation. The SLLOD method generates the velocity profile for Couette flow. If we choose the flow along the $x$-axis and the gradient of the flow along the $y$-axis, the shear rate dependent viscosity is given by
\begin{equation}
\eta = - \frac{\langle P_{\rm xy} \rangle}{\dot \gamma} 
\label{Equation:Viscosity}
\end{equation}
where $\langle P_{\rm xy} \rangle$ is the average of the $x$-$y$ components of the pressure tensor (sometimes also called the stress tensor), and $\dot \gamma$ is the shear rate~\cite{Allen_Tildesley}. We limit our simulations to small enough shear rates to avoid potential instabilities associated with breaking FENE bonds in the simulation~\cite{Xu_DePablo_Kim}.

We choose a loading fraction $\phi \approx 0.05$, defined by the ratio of the number of nanoparticle force sites to the total number of system force sites; values of this order of magnitude are common in experimental studies.~\cite{Ren_Krishnamoorti,Hsied,Kumar,Xu} Since all force sites have the same diameter, $\phi$ should also be roughly equal to the volume fraction. Due to the fact both polymers and nanoparticles consist of a discrete number of force sites, $\phi$ varies slightly between systems. For the icosahedral systems $\phi = 0.0494$, and for the rods and sheets $\phi = 0.0505$. The system size for all three nanoparticle systems is much larger than either the radius of gyration of the polymers, or the largest nanoparticle dimension, to avoid finite size effects. The number of chains $N_{\rm C}$ ranges from $N_{\rm C}=10$ in the smallest system up to $N_{\rm C}=994$ for the largest systems. For each system we examine chain lengths $N=10$, $20$, and $40$. For this polymer model, the entanglement length $\approx$ 30 to 40~\cite{Putz_Kremer_Grest}. Hence, only the longest chains studied may exhibit any effects of entanglement and the effects should be small in the present work. We reiterate that effects of dynamics due to proximity to the glass transition should not play a significant role at the relative high $T$ of our systems. A summary of the system parameters can be found in Table~\ref{Table1}.

\section{Composite Rheology}
\label{sec:Composite_Rheology}

\subsection{Shear Viscosity $\eta$}
\label{sub:Composite_Rheology_Viscosity}

In this section, we evaluate the role of nanoparticle geometry on the magnitude of $\eta$ and clarify the influence of chain bridging on $\eta$ for a range of shear rates ($\dot \gamma = 0.005$, 0.007, 0.01, and 0.02) and chain lengths. In the nanocomposite simulations of~\cite{Starr_Douglas_Glotzer}, $\eta$ became nearly constant at $\dot\gamma =0.005$ -- in other words, we are near the Newtonian limit where $\eta$ approaches a stationary value for $\dot\gamma \rightarrow 0$. Thus, we expect our lowest shear rates approach the Newtonian regime; in this sense, the shear rates we use are fairly low. However, if we consider $\dot\gamma$ in physical units, these rates would be extremely high by experimental standards. Ideally, we would estimate $\eta$ in the $\dot\gamma = 0$ limit directly (for example, by using an Einstein or Green-Kubo relation), but the  accurate evaluation of $\eta$ using these methods is difficult~\cite{Smith_Bedrov_Li_Byutner,Sen_Kumar_Keblinski}.

\begin{figure}[t]
\begin{center}
\includegraphics[clip,width=8.6cm]{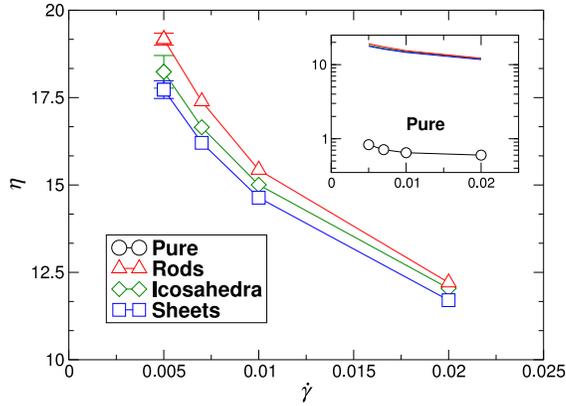}
\end{center}
\caption{The viscosity $\eta$ as a function of shear rate $\dot \gamma$ for chain length $N = 20$ at $\rho=1.00$. The rods show the largest $\eta$, followed by the icosahedra, and lastly the square sheets. We show the statistical uncertainty of $\eta$ for each system at $\dot \gamma = 0.005$, where the fluctuations is largest, and hence represents an upper bound for the relative uncertainty for all $\eta$ calculations. The uncertainty is the result of ``block averaging''~\cite{Allen_Tildesley}. The inset includes the bulk polymer and shows that the pure system has much lower viscosity than any of the nanocomposites by roughly an order of magnitude. The lines in this and subsequent figures are drawn only as guide for the reader's eyes.}
\label{fig:Shear_Figure_1}
\end{figure}

\begin{figure}[t]
\begin{center}
\includegraphics[clip,width=8.6cm]{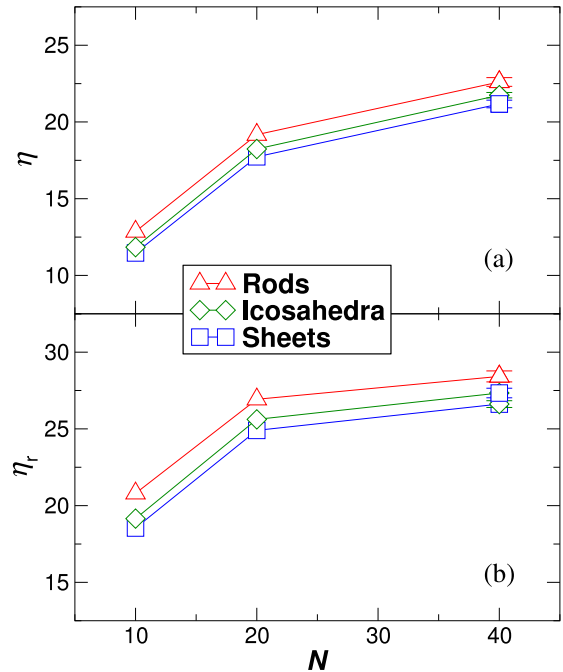}
\end{center}
\caption{The (a) viscosity $\eta$ and (b) reduced viscosity $\eta_{\rm r}$ as a function of chain length $N$. All values are from systems at $\rho=1.00$ and $\dot \gamma=0.007$. Chain length appears to have no effect on the ordering of $\eta$ amongst the systems.}
\label{fig:Shear_Figure_2}
\end{figure}

Figure~\ref{fig:Shear_Figure_1} shows $\eta$ for the three nanocomposites and pure polymer as a function of $\dot \gamma$. These data show a clear change in $\eta$ between the systems for all $\dot \gamma$ simulated, with the rods having the largest $\eta$, followed by the icosahedra, sheets, and finally the bulk polymer. The decrease of $\eta$ with increasing $\dot \gamma$ is indicative of shear thinning. We note that $\eta$ is not constant at the lowest $\dot \gamma$, indicating we have not yet reached the Newtonian regime. Nonetheless, it is clear that the pure polymer has a viscosity approximately an order of magnitude smaller than any of the composites. Comparable differences in $\eta$ have been observed in experiments~\cite{Ray_Okamoto,Potschke_Fornes_Paul}. This demonstrates an enhancement of $\eta$ through the addition of nanoparticles, which is often a desired goal of adding nanoparticles to a polymer melt. While the figure only shows $\eta$ for the $N=20$ and $\rho=1.00$, other chain lengths and densities indicate the same trend. Figure~\ref{fig:Shear_Figure_2} (a) shows the effect of varying $N$ on $\eta$ at $\rho=1.00$ and $\dot \gamma=0.007$. An increase in $\eta$ with increasing $N$ is expected from basic polymer physics since the chain friction coefficient increases with $N$~\cite{Doi_Edwards}. Interchain interactions and ``entanglement'' interactions enhance this rate of increase since the friction coefficient of each chain increases linearly with $N$. 

While it is clear that addition of nanoparticles increases $\eta$ of the resulting composite, the reasons for this are not obvious. Since the pure polymer has a relatively low $\eta$ it is reasonable to assume that the rigid nanoparticles themselves inherently raise $\eta$ as in any suspension of particles in fluid matrix. To separate out this continuum hydrodynamic effect of the nanoparticles from the effects specifically due to polymer-nanoparticle interactions, we calculate intrinsic viscosity $[\eta]$ for the individual nanoparticles. Specifically, continuum hydrodynamics provides an estimate of the incremental change of $\eta_{\rm r}$ through the addition of particles to a fluid, 
\begin{equation}
\eta_{\rm r}(\phi) = 1 + [\eta] \phi + \mathcal{O}(\phi^2)
\label{Equation:Intrinsic_Viscosity}
\end{equation}
where the velocity of the fluid is taken to be zero on the surface of the particle (``stick'' boundary conditions). Such expansions are only quantitatively reliable for $\phi \lesssim 0.01$ for nearly spherical particles; for larger $\phi$, we expect eq.~(\ref{Equation:Intrinsic_Viscosity}) to recover only qualitative variations in $\eta_r$.   For reference, $[\eta] = 5/2$ for spheres in three dimensions~\cite{Einstein_Ford}. It is also known that $[\eta]$ is generally larger for isotropically oriented spherical shapes~\cite{Douglas_Garboczi}. Refs.~\cite{Bicerano_Douglas_Brune,Surve_Pryamitsyn_Ganesan,Ganesan_Macromolecules,Ganesan_Journal_of_Rheology} have reviewed the modeling of polymer composites containing extended particles based on continuum mechanics. This work also considers the effect of particle clustering on $\eta$ which is important under poor dispersion conditions. 

To compare the predictions of hydrodynamic theory to our simulation, we use the Zeno package to calculate $[\eta]$ for our particles~\cite{Zeno_FN}. The computational method involves enclosing an arbitrary-shaped object within a sphere and launching random walks from the border or launch surface. The probing trajectories either hit the object or return to the launch surface. From these path integration calculations, $[\eta]$ for the rigid particles can be estimated for objects of essentially arbitrary shape~\cite{Mansfield_Douglas_Garboczi}. We determine $[\eta]$ of the nanoparticles as an average over $100$ different confirmations found in the equilibrium system as they adopt a range of shapes due to bond flexibility.

Using Zeno, we find that the icosahedral nanoparticles have $[\eta] = 3.93$, the rods have $[\eta] = 16.3$, and the sheets have $[\eta] = 10.8$. The intrinsic viscosity of a regular icosahedron has been estimated to be 2.47~\cite{Mansfield_Douglas_Garboczi}. The discrepancy between this value and our icosahedral nanoparticles arises from the fact that our nanoparticles are not regular solids, but rather consist of a group of spherical force sites with icosahedral symmetry. We can use Eq.~(\ref{Equation:Intrinsic_Viscosity}) to predict $\eta_{\rm r}$ for the MD simulations, and find (for the $N=20$ systems) $\eta_{\rm r}= 1.20$ for the icosahedra, $\eta_{\rm r}= 1.82$ for the rods, and $\eta_{\rm r}= 1.54$ for the sheets. Such a prediction underestimates the $N=20$ MD results by $92~\%$ for the icosahedra, $90~\%$ for the rods, and $89~\%$ for the sheets. Clearly this hydrodynamic estimate of $\eta_{\rm r}$ for our nanocomposites is not adequate, even accounting for the order-of-magnitude nature of the estimate based on eq.~(\ref{Equation:Intrinsic_Viscosity}). Consistent with the full MD simulations, the rods have the largest value, but in contrast to the MD simulation, the order of the sheets and icosahedra are reversed. While the larger $[\eta]$ value of the rods are generally consistent with a larger value of $\eta$, the reversal of the sheet and icosahedra $[\eta]$ values indicates that nanoparticle and/or polymer interactions play a role in the overall viscosity. Moreover, the continuum theory indicates that $\eta_{\rm r}$ should be {\it independent} of $N$, while we see from Fig.~\ref{fig:Shear_Figure_2} (b) that $\eta_{\rm r}$ strongly depends on $N$. Hence we conclude that the continuum model of $\eta$ does not provide an adequate explanation of the changes we see in our simulations.

Evidently, we must also examine other contributions to the nanoparticle viscosity. One possibility that has recently been discussed~\cite{Ganesan_Journal_of_Rheology} is that slip rather than stick boundary conditions might provide a better description of nanoparticle boundary conditions. However, this would result in a {\it decrease} of the predicted value of $\eta_{\rm r}$~\cite{Ganesan_Pryamitsyn_Surve_Narayanan,Smith_Bedrov_Li_Byutner} relative to the stick case, which is in the wrong direction for explaining our results. At a molecular level, the boundary conditions are clearly neither stick nor slip, as particles have finite residence times at the surface, although it is difficult to determine hydrodynamic boundary conditions directly from molecular considerations.  Nanoparticle-polymer interactions and nanoparticle clustering are evidently factors that might cause the property enhancements we observe, and in the next section we focus on this possibility.

\subsection{Relationship of Viscosity to Fluid Structure}
\label{sub:Composite_Rheology_Structure}

We next turn to the structural results obtained from MD simulation to better quantify the relationship of polymer and nanoparticle structure to $\eta$. We first calculate the fraction of nanoparticles in contact with other nanoparticles, $f_{\rm N-N}$. Nanoparticles are said to be ``in contact'' if one or more comprising force sites of a nanoparticle are within the nearest neighbor distance of a force site belonging to another nanoparticle. The nearest-neighbor distance is defined by the first minimum in the radial particle distribution function $g(r)$; the first minima is at $r \approx 1.5$ for all systems. Figure~\ref{fig:NP-NP} shows $f_{\rm N-N}$ as a function of $N$ for $\rho = 1.00$ and for both $\dot \gamma = 0$ and $\dot \gamma = 0.007$. While it is not readily apparent from Fig.~\ref{fig:Systems} (c), the polymers ``intercalate'' between the sheets to some degree, and hence, the nanoparticles are not actually in direct contract with each other, resulting in a very low value of $f_{\rm N-N}$ for the sheets. With the exception of the $N = 40$ equilibrium system, the sheets have the fewest nanoparticle-nanoparticle contacts. For the systems under steady shear, $f_{\rm N-N}$ evidently stays fairly constant as $N$ varies. While the order of the systems for $f_{\rm N-N}$ is consistent with the trend for $\eta$, the similarity between the rods and icosahedra suggests that the fraction of nanoparticles alone is not responsible for determining composite viscosity. Of note, is the large fraction of nanoparticle contacts for the $N = 40$ sheet system at equilibrium. More in-depth analysis shows that the sheets favor a clustered state at lower density, i.e. chains no longer fully intercalate between the sheets. Both the $N = 10$ and $N= 20$ sheet systems have intercalating polymers, while the $N = 40$ case does not. The tendency of the sheets to stack is due in part to entropic interactions that result from depletion attractions between the sheet; the depletion interactions arise due to physically adsorbed polymers on the sheet surface, much like the depletion interactions of polymer coated colloids.~\cite{Anderson_Lekkerkerker} This interaction is most pronounced for the sheets, presumably due to the large and relatively flat surface of a sheet, which more effectively reduces the entropy of chains at the surface.


Given that nanoparticle-nanoparticle contacts do not appear to be the origin of the relative ordering, we next consider the role of polymer-nanoparticle interactions. To gauge the possible role of such interactions we calculate the fraction of polymer chains in direct contact with a nanoparticle, $f_{\rm N-P}$. Using the same nearest-neighbor criteria as used to determine nanoparticle-nanoparticle contacts, we compute $f_{\rm N-P}$ as a function of $N$ for $\rho=1.00$ at $\dot \gamma = 0$ and $\dot \gamma = 0.007$ in Fig.~\ref{fig:Fraction}. We find that $f_{\rm N-P}$ is approximately the same for both the quiescent and sheared systems. There are also clear trends: $f_{\rm N-P}$ increases with increasing $N$ and for every value of $N$ the system containing the rods have the largest value of $f_{\rm N-P}$, followed by the system containing the icosahedra, and finally the system containing the sheets. These same trends in $\eta$ appear in Fig.~\ref{fig:Shear_Figure_2} suggesting that the nanoparticle-polymer contacts indeed have a significant effect on relative magnitude of $\eta$. 

\begin{figure}[t]
\begin{center}
\includegraphics[clip,width=8.6cm]{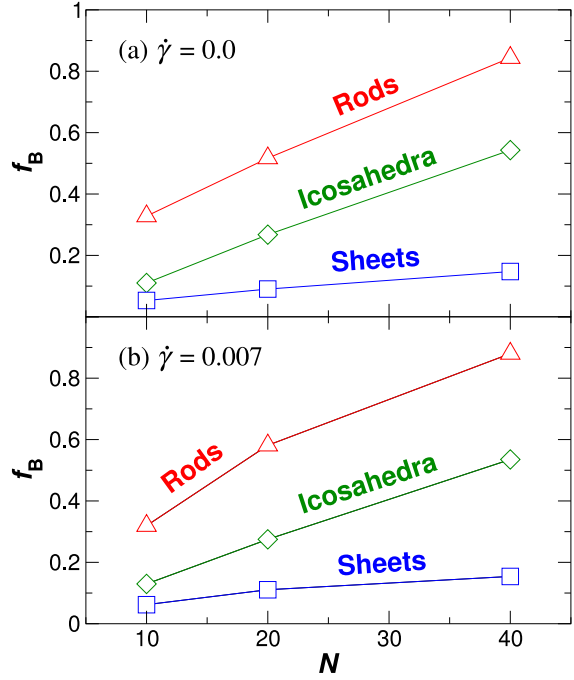}
\end{center}
\caption{The effect of chainlength $N$ on the fraction of bridging chains $f_{\rm B}$ in each of the three nanoparticle systems at $\rho = 1.00$ for (a) $\dot \gamma=0$ and (b) $\dot \gamma=0.007$. The order of $f_{\rm B}$ with respect to nanoparticle type matches the trend in Fig.~\ref{fig:Shear_Figure_1} for $\eta$.}
\label{fig:Bridging}
\end{figure}

The correlation of $\eta$ with $f_{\rm N-P}$ is consistent with the idea that bridging between nanoparticles play a major role in determining the rheological properties of polymer nanocomposites~\cite{Salaniwal_Kumar_Douglas,Hooper_Schweizer_2005,Hooper_Schweizer_2006,Gersappe}. Thus, we extend our analysis to test for a correlation between $\eta$ and chain bridging. We define a ``bridging chain'' as a chain that is simultaneously in contact with two or more nanoparticles. Figure~\ref{fig:Bridging} shows $f_{\rm B}$ as a function of $N$ for $\rho=1.00$. The trend in $f_{\rm B}$ is consistent with both $f_{N-P}$ and $\eta$, showing an increase in $f_{B}$ with increasing $N$ as well as a clear ordering between systems for every chain length. Although not shown, the results seen in Fig.~\ref{fig:Bridging} occur for every value of $\rho$ that we have simulated. Hence, the fraction of bridging chains seems to be a useful ``order parameter'' for characterizing the polymer-nanoparticle interactions. If the polymer-nanoparticle interactions were sufficiently small, the bridging would not be expected to play significant role, and it is likely that the continuum hydrodynamic approach would be applicable.

\begin{figure}[t]
\begin{center}
\includegraphics[clip,width=8.6cm]{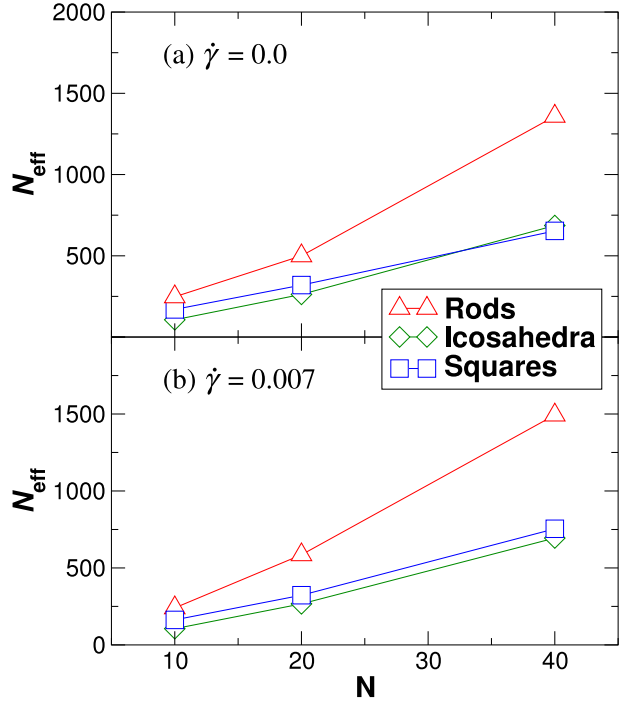}
\end{center}
\caption{The effective chainlength $N_{\rm eff}$ as a function of $N$ in each of the three composite systems at $\rho = 1.00$ for (a) $\dot \gamma=0$ and (b) $\dot \gamma=0.007$.}
\label{fig:Effective_Chains}
\end{figure}

Similar to the notion of bridging is the idea that the nanoparticles can act as transient cross-linkers that can lead to effects equivalent with the formation of higher molecular weight chains~\cite{Gersappe,Salaniwal_Kumar_Douglas}. To test this idea, we define an ``effective chain'' as the collection of chains that are connected by nanoparticles. We then define an effective chain length $N_{\rm eff}$ as the mean mass of chains connected by the nanoparticles. Fig.~\ref{fig:Effective_Chains} shows that $N_{\rm eff}$ is almost an order of magnitude larger than $N$. Hence, we expect that the largest contribution to the $\eta$ increase comes from this effect. Even in the simplest Rouse theory, an order of magnitude increase in $N$ would lead to an order of magnitude increase in $\eta$. The formation of longer effective chains is expected to lead to entanglement interactions that would further amplify the $\eta$ increase, as emphasize by ref.~\cite{visc-entang}. However, this entanglement contribution is hard to quantitatively interpret in the present context.

\section{Isotropic Tensile Strength $\tau$}
\label{sec:Isotropic_Tensile_Strength}

\subsection{Calculation of $\tau$}
\label{sub:Isotropic_Tensile_Strength_Calculation}

The ultimate isotropic tensile strength $\tau$ of a material is defined as the maximum tension a homogeneously stretched material can sustain before fracture. While this definition can be directly probed experimentally, the situation is less straightforward and computationally accessible in an MD simulation. Ref.~\cite{Shen_Debenedetti_Stillinger} has developed an approach to estimate $\tau$ based on potential energy-landscape (PEL) formalism that is accessible by MD simulation. Although the PEL is complex and multidimensional, one can envision it as a series of energy minima connected by higher energy transition pathways. By definition, the minima, or inherent structures (IS) are mechanically stable, i.e. there is no net force at the minimum. The method of ref.~\cite{Shen_Debenedetti_Stillinger} relates $\tau$ to the maximum tension the IS can sustain, determined by an appropriate mapping from equilibrium configurations on the PEL. In other words, $\tau$ is the upper bound on the tension at the breaking point of an ideal glass state in the $T\rightarrow 0$ limit. Here we evaluate $\tau$ for the various possible nanoparticle geometries. As a cautionary note, we point out that this method has not been directly validated by comparison with experiments; thus conclusions drawn from these results should be considered tentative. Ref.~\cite{Kopsias_Theodorou} discusses an alternate approach using the PEL to evaluate the elastic constants. 

To determine $\tau$, we must first generate the energy minimized configurations from the equilibrium configurations. For a system of $N$ atoms in the canonical ensemble, the most commonly used mapping from equilibrium configurations to the IS takes each force site along the steepest decent path of the energy of the system. This configurational mapping procedure corresponds to the physical process of instantaneous cooling to the $T\rightarrow0$ limit to obtain an ideal glass with no kinetic energy. Therefore, by sampling thermally equilibrated configurations and minimizing their energies we can estimate the average inherent structure pressure, $P_{\rm IS}$, which will be negative when the system is under tension. The maximum tension then defines $\tau$. The value of $P_{\rm IS}$ has been found to be weakly dependent on the $T$ at which the sampling is performed. For example, while cyclopentane has a strongly temperature dependent structure, ref.~\cite{Utz_Debenedetti_Stillinger} has shown that equilibrium $T$ only weakly effects $P_{\rm IS}$. Since the structures in our systems do not display such dependence, we expect our results also to be independent of the starting equilibrium $T$. Thus, we can use the same systems at $T=2$ that we used to determine $\eta$ and still obtain reliable results for $\tau$ -- even though the starting configuration is a highly fluid state. This calculation of $\tau$ should be an {\it upper bound} for the tensile strength that would be obtained for any finite $T$ system, and hence is referred to the as the ultimate tensile strength. The energy minimization process eliminates the high frequency effects that would normally be associated with such high $T$ states and thus the method refers to the strength of an ideal glass material having essentially no configurational entropy. 

\begin{figure}[t]
\begin{center}
\includegraphics[clip,width=8.6cm]{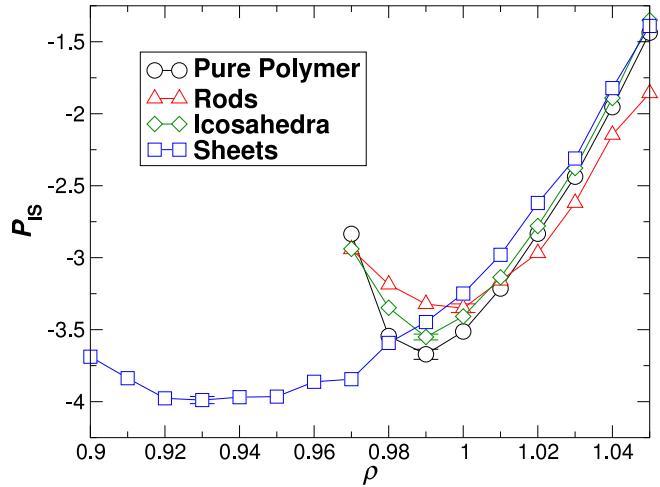}
\end{center}
\caption{The inherent structure pressure $P_{\rm IS}$ as a function of $\rho$ for the pure polymer and three nanocomposite systems at $N=20$. The pure, icosahedron, and rod systems all have a Sastry density $\rho_{\rm S} \approx 0.99$, while for the sheets $\rho_{\rm S} \approx 0.93$. The uncertainty intervals (``error bars'') represent the statistical uncertainty in our average for $P_{IS}$ from block averaging~\cite{Allen_Tildesley}. }
\label{fig:Congrad_N20}
\end{figure}

Figure~\ref{fig:Congrad_N20} shows the $P_{\rm IS}$ curves generated for the $N = 20$ variant of each system. We find that $P_{\rm IS}$ decreases at large $\rho$, until a minimum value is reached. The density of the minimum, referred to as the ``Sastry density'' $\rho_{\rm S}$~\cite{Sastry_Debendetti_Stillinger}, characterizes the density where the system fractures, and voids first start to appear in the minimized configurations. As $\rho$ decreases below this point, $P_{\rm IS}$ begins to rise. Since this is the greatest tension achievable, we have $P_{\rm IS}^{\rm MAX} = -\tau$. For pure, icosahedral, and rod systems we find $\rho_{\rm S} \approx 0.99$, while for the sheets it is much lower $\rho_{\rm S} \approx 0.92$. Similarly, $\tau$ is considerably larger for the sheets than for the other systems. We find that the increase in $\tau$ for the sheets is not as sizable as observed experimentally~\cite{Ray_Okamoto}. This may be related to the fact that (experimentally) the chains often form stronger associations with layered silicates than in our simulations; additionally, the chain lengths we examine are small compared to experiments, and we will see that $\tau$ increases with $N$. We note that at this chain length, the icosahedra and rods actually {\it reduce} the strength of the nanocomposite as compared to the pure melt. 

\begin{figure}[t]
\begin{center}
\includegraphics[clip,width=8.6cm]{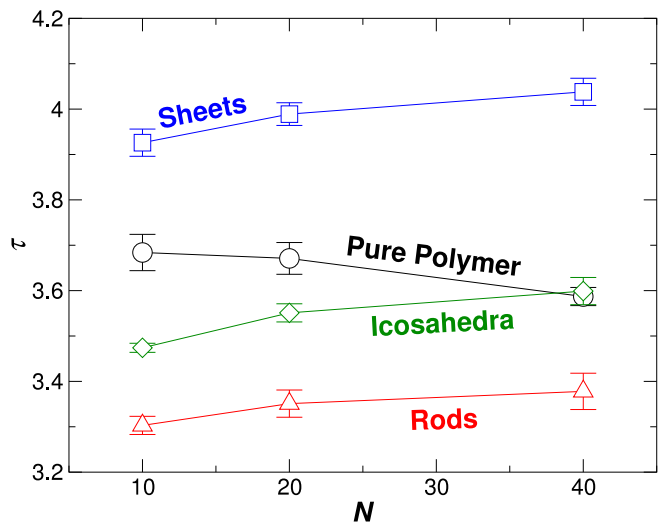}
\end{center}
\caption{Tensile strength $\tau$ as a function of chain length $N$ for the pure polymer and three nanocomposite systems. We find that as $\tau$ increases as $N$ increases. This is in contrast to the behavior of the pure polymer where $\tau$ decreases with increasing $N$. Only for systems with the sheet nanoparticles --- or the system with icosahedral nanoparticles at $N\ge40$ --- does the addition of nanoparticles produce a net benefit relative to the bulk polymer.}
\label{fig:Total}
\end{figure}

To quantify the chain length dependence of $\tau$, we plot the minimum of each $P_{\rm IS}$ curve as a function of $N$ in Fig.~\ref{fig:Total}. For the pure polymers, we find that $\tau$ {\it decreases} with increasing chain length. This is non-trivial, since one might na\"{i}vely expect longer chains to exhibit more interchain coupling. This is the same trend with chain length as observed in ref.~\cite{Shen_Debenedetti_Stillinger} for $n$-alkanes. It is reassuring that simple bead-spring model leads to similar results but this does not help us build intuition about the physical meaning of these results. Reference~\cite{Shen_Debenedetti_Stillinger} finds a maximum $\tau$ for $N=3$, and a tendency for $\rho_{\rm IS}$ to saturate near $N=8$. Since the smallest chain we simulate is $N=10$, we are above the regime where this feature occurs. Consistent with these facts, $\tau$ decreases with $N$ and $\rho_{\rm S}$ is roughly independent of $N$. 

Most importantly, Fig.~\ref{fig:Total} shows that the addition of the nanoparticles {\it reverses} the $N$ dependence of $\tau$ when compared to the pure system. Thus, while the presence of icosahedral or rod-like nanoparticles decreases the material strength for most chain lengths studied, the trend of $\tau$ is increasing, and if this continues, $\tau$ will surpass the pure melt for all nanoparticle shapes at large enough $N$. Indeed, $\tau$ for the icosahedra nanocomposite already exceeds that of the melt at $N=40$. 

\subsection{Relationship of Structure to Tensile Strength} 
\label{sub:Isotropic_Tensile_Strength_Structure}

\begin{figure}[t]
\begin{center}
\includegraphics[clip,width=8.6cm]{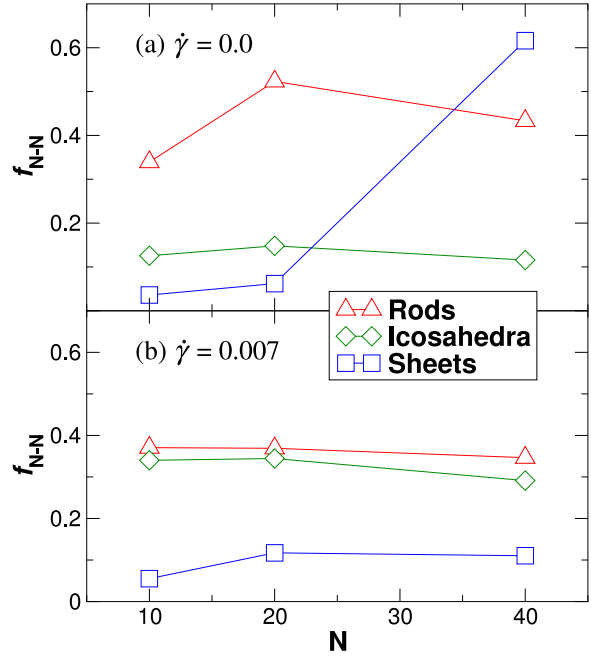}
\end{center}
\caption{The fraction of nanoparticles in contact with other nanoparticles $f_{\rm N-N}$ as a function of $N$ for each of the composite systems at (a) $\dot \gamma=0$ and (b) $\dot \gamma=0.007$.}
\label{fig:NP-NP}
\end{figure}

\begin{figure}[t]
\begin{center}
\includegraphics[clip,width=8.6cm]{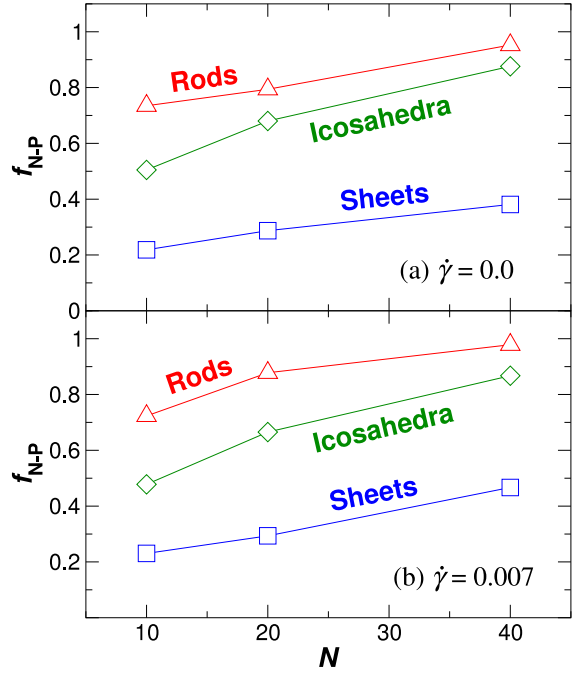}
\end{center}
\caption{The fraction of polymer chains in direct contact with a nanoparticle $f_{\rm N-P}$ shown as function of chain length $N$ for each of the three nanoparticle systems at (a) $\dot \gamma=0$ and (b) $\dot \gamma=0.007$.}
\label{fig:Fraction}
\end{figure}

To better understand the predicted $N$ dependence of $\tau$, we perform a parallel analysis to the structural analysis of shear runs discussed above. The analysis of Figs.~\ref{fig:NP-NP}-\ref{fig:Bridging} focused on $\rho=1.00$. We relate the structure to $\tau$ by examining the structure at $\rho_{\rm S}$. Evidently, $\rho_{\rm S}$ differs for the sheets in relation to the other systems studied, and it is possible that $\rho$ dependent changes in connectivity properties are responsible for the difference of $\tau$. Thus, we calculate the quantities $f_{\rm B}$, $f_{\rm N-N}$, and $f_{\rm N-P}$ at $\rho_{\rm S}$ for each system using the equilibrium and IS configurations at the Sastry density $\rho_{\rm S}$ of the system. The results for the equilibrium and IS configurations show the same qualitative trends, so we will present only data for the IS. 

\begin{figure}[t]
\begin{center}
\includegraphics[clip,width=8.6cm]{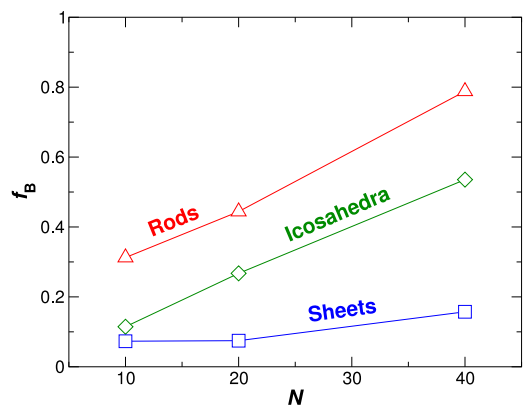}
\end{center}
\caption{The effect of chainlength $N$ on the fraction of bridging chains $f_{\rm B}$ in each of the three composite systems calculated using the inherent structure configurations at $\rho = \rho_{\rm S}$.}
\label{fig:Bridging_Minimized}
\end{figure}

We first focus on $f_{\rm B}$ because our investigations into the rheological properties suggested that bridging chains played a large role in determining $\eta$. However, when we plot $f_{\rm B}$ in Fig.~\ref{fig:Bridging_Minimized} for each system at $\rho_{\rm S}$, we see that the $N$ dependence is the {\it reverse} of that seen for $\tau$, even though this quantity evidently follows $\eta$. Even for the sheets which have a $\rho_{\rm S}$ significantly lower value of than the other systems, the ordering of $f_{\rm B}$ among the systems is the same as for $f_{\rm B}$ in the case of the $\eta$ calculations. The sheet composites have the smallest $f_{\rm B}$, followed by those with icosahedra, and those with rods. While bridging chains do increase with increasing $N$, $f_{\rm B}$ does {\it not} seem to be a major factor in the relative value of $\tau$ between the composites. For example, the $N=10$ nanocomposites with icosahedra and sheets have almost the same value of $f_{\rm B}$, yet they have significantly different values for $\tau$. 

\begin{figure}[t]
\begin{center}
\includegraphics[clip,width=8.6cm]{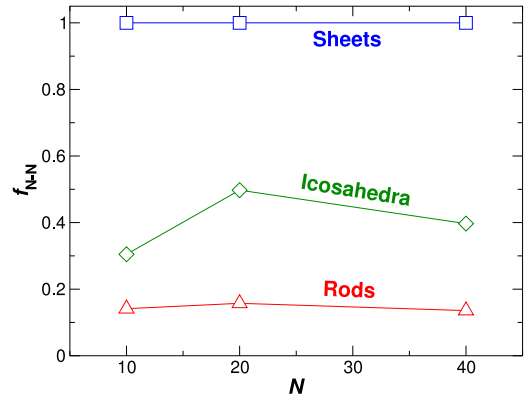}
\end{center}
\caption{The fraction of nanoparticles in contact with other nanoparticles $f_{\rm N-N}$ as a function of $N$ for each of the composite systems calculated using the inherent structure configurations at $\rho = \rho_{\rm S}$. }
\label{fig:NP-NP_Minimized}
\end{figure}

The fewer bridging chains in the systems with square sheets suggests that the sheets may be clustered, and we find that this is indeed the case for $N=40$ (Fig.~\ref{fig:NP-NP}~(a)). Thus, we plot $f_{\rm N-N}$ in Fig.~\ref{fig:NP-NP_Minimized} and find indeed that $f_{\rm N-N} \approx 1$ for the sheets. The sheets prefer a stacked state at low $\rho$ due to entropic interactions with polymers which leads to a reduced explicit energetic interaction with the surrounding polymers. In particular, the relative ordering of $f_{\rm N-N}$ matches that of $\tau$. Na\"{i}vely, this would seem to suggest a potential correlation between $f_{\rm N-N}$ and $\tau$. However, this apparent correlation is problematic. Firstly, the $N$ dependence of $\tau$ and $f_{\rm N-N}$ are different, namely $\tau$ monotonically increases, while $f_{\rm N-N}$ is roughly constant. Secondly, if the nanoparticles interactions are the origin of the increased strength, then one might expect to find the largest  effect when the nanoparticles' are well dispersed~\cite{Starr_Douglas_Glotzer}. This is not the case, since the clustered sheets give the {\it largest} effect.

Ref.~\cite{Shen_Debenedetti_Stillinger} demonstrated that the rupture of the system occurs in ``weak spots.'' To confirm that the nanoparticles actually impart additional strength, we visually examined the location of fracture in our systems at $\rho_{\rm S}$. To find empty spaces that are at least the size of a particle, we discretize the system into a cubic lattice of overlapping spheres, each with a radius $r_{\rm V}$ and a nearest neighbor separation $d$. We then identify spheres that do not contain any system force site. By adjusting the parameters $r_{\rm V}$ and $d$ we can ensure that the voids we find are at least large enough for a single particle. Since the force sites of both the nanoparticles and polymers have diameter $1$, we must choose $r_{\rm V} > 0.5$ to have physical relevance. We find that values of $r_{\rm V} = 0.6$ and $d=0.3$ work best for visualizing voids. 

\begin{figure}[t]
\begin{center}
\includegraphics[clip,width=8.6cm]{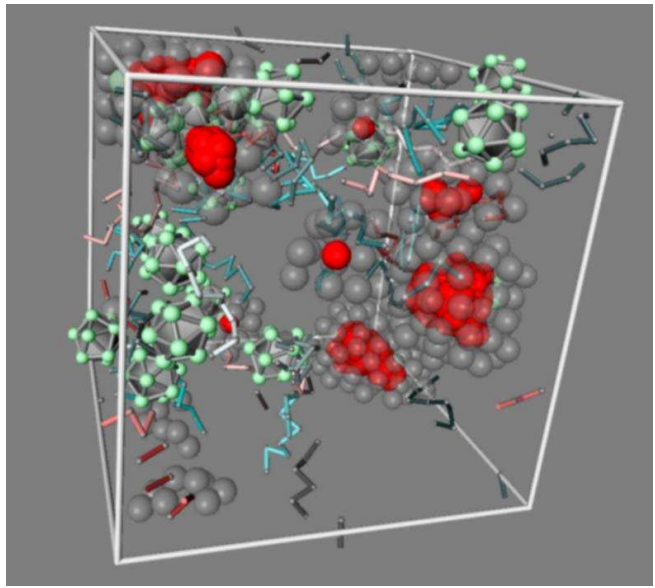}
\end{center}
\caption{Typical fractures for the IS of the icosahedral nanocomposite. The large contiguous blobs are the fractures. The surrounding spheres represent force sites in contact with the rupture. Note that nearly all sites in contact are polymers. }
\label{fig:Voids}
\end{figure}

Figure~\ref{fig:Voids} shows that the voids in the system occur in regions of pure polymer for the icosahedral system. The same is also true for the rods and square sheets. Quantitative analysis of the force sites within nearest-neighbor distance $r=1.5$ of the voids confirmed this suggestion, with over 99~\% of the resulting force sites belonging to polymer chains. Thus, while the nanoparticles impart an increased strength to the nanocomposite, the correlation to $f_{\rm N-N}$ is not apparently causal. Evidently, there is something more subtle controlling the magnitude of $\tau$ in the sheet filled nanocomposite that we have not yet identified.

\begin{figure}[t]
\begin{center}
\includegraphics[clip,width=8.6cm]{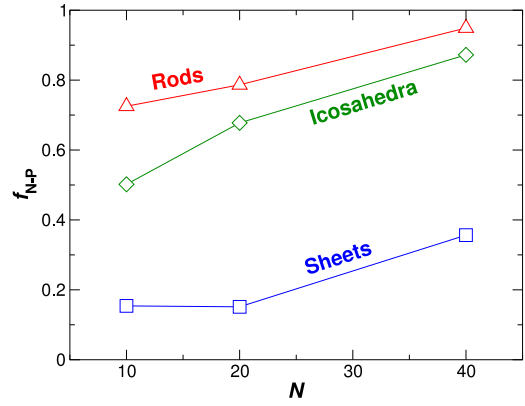}
\end{center}
\caption{The fraction of polymer chains in direct contact with a nanoparticle $f_{\rm N-P}$ as function of chainlength $N$ for each of the three nanoparticle systems calculated using the inherent structure configurations at $\rho = \rho_{\rm S}$.}
\label{fig:Fraction_Minimized}
\end{figure}

To complete the structural analysis we calculate $f_{\rm N-P}$ for the IS. Figure~\ref{fig:Fraction_Minimized} shows that the sheet nanoparticles are not in contact with many polymers, consistent with the fact that $f_{\rm N-N} \approx 1$. In fact, after minimization the difference in $f_{\rm N-P}$ between the rods and the sheets is even larger than observed in the equilibrium configurations shown in Fig.~\ref{fig:Fraction}. Figure~\ref{fig:Fraction_Minimized} shows that the relative ordering of $f_{\rm N-P}$ is {\it opposite} to that of $\tau$, hence the number of nanoparticle-polymer contacts alone also does not provide a good indicator of $\tau$.

A potential clue to the increase of $\tau$ of the sheet nanocomposite is the fact that $\rho_{\rm S}$ is {\it significantly smaller} than for the other systems. This allows the sheet composites to undergo a greater deformation before fracture. This large increase in both the strength and toughness of the polymer matrix with incorporation of nanoparticles is reminiscent of the changes in the properties of natural and synthetic rubbers with the inclusion of carbon black and nanofiller additives~\cite{Medalia,Kenny_McBrierty_Rigbi_Douglas,Rharbi,Gersappe}. Extraordinarily large increases in both the strength and toughness of materials have recently been observed in the case of exfoliated clay sheets dispersed in the polymer polyvinylidene fluoride (PVDF)~\cite{Shah}. Although some of this change is associated with the modification of the crystallization morphology by the clay nanoparticles, this does not explain in itself the observed toughening mechanism. It is known that nanofiller particles can behave as temporary cross-linking agents that impart viscoelastic characteristics to the fluid. Gersappe~\cite{Gersappe} further emphasizes the role of this transient network formation in impeding cavitation events that initiate material rupture and the potential significance of this phenomenon in understanding the nature of biological adhesives (abalone), fibers (spider silk), and shell material (nacre)~\cite{Smith,Tsagaropoulos_Eisenberg}. Gersappe further suggests that the relative mobility of nanoparticles has a role on the toughening of materials. However, such reasoning is inconsistent with the fact that the sheet nanoparticles lead to the strongest material while being the {\it least mobile} of our additives.

Recent work indicates that flexible sheet-like structures, such as studied here, are characterized by a {\it negative} Poisson ratio $\nu$~\cite{Boal_Seifert_Shillcock,Bouick,Lakes}, {\it i.e.}\ the material expands normal the the direction of stretching, as opposed to normal materials which expand in the same direction as the stretching. The formation of a composite material with negative Poisson ratio (``auxetic'' materials) is expected to lead to a reduction of the Poisson ratio of the composite as a whole~\cite{Wei_Edwards,Garboczi_Douglas_Bahn}. A large reduction in $\nu$ is expected to give rise to materials that are strong and fracture resistant~\cite{Choi_Lakes}. Related to this, we observe in our simulations that the voids that form in the sheet-filled nanocomposites tend to be fewer and smaller than those for the other nanoparticles at $\rho_{\rm S}$. As we further reduce $\rho$ for the sheet system, the voids grow more numerous as opposed to growing larger as in the other nanocomposites. Microvoid formation has also been established as a mechanism for toughening polymer materials~\cite{Bagheri}. Previous work devoted to understanding the high impact strength of polycarbonate and other glassy polymers has likewise emphasized the importance of large ``free volume'' within the polymer material as a necessary condition for large toughness~\cite{Mercier_Aklonis_Tobalsky,Matsuoka_Ishida}. Based on our simulation results and previous observations, we tentatively suggest that the addition of these ``springy'' sheet materials reduces the effective Poisson ratio of our nanocomposites, and that the microvoid formation process that we observe is a manifestation of the non-uniformities in the elastic constants within the nanocomposite. In principle, the elastic constants can be determined by the PEL approach~\cite{Kopsias_Theodorou}, but this analysis will be deferred to a future work since it is rather involved. Our tentative interpretation of the predicted $\tau$ variation in the sheet nanocomposites also suggests the need for better characterization of the geometric rigidity properties of the sheet nanoparticles, since these variables may be important for understating how such particles modify the stiffness and toughness of nanocomposites.

\section{Conclusion}
\label{sec:Conclusion}

In this work we have focused on how nanoparticle shape influences the viscosity of polymer-nanoparticle melt mixtures at high temperature and the ultimate tensile strength (estimated from the PEL formalism) of polymer nanocomposites in the ideal glass state. Our results suggest that chain bridging between the nanoparticles can have a large effect on $\eta$ of the mixture when the polymer-particle interactions are attractive, so that the nanoparticles disperse readily.  
In addition, there is a relatively weak increase in $\eta$ for sheet nanoparticle composites, which tend to cluster in our simulations, supporting the expectation that nanoparticle clustering diminishes the viscosity enhancement. (However, the formation of ``open'' or percolating fractal clusters may have the opposite effect on viscosity). Curiously, the tensile strength of the sheet nanocomposite is greatest, in spite of the sheet stacking. One of the most intriguing effects of the nanoparticles is that, regardless of shape, the dependence of the tensile strength on chain length for the nanocomposites is {\it opposite} to that for the pure polymer melt. We reiterate that our results for $\tau$ rely on the PEL approach, and this approach should be carefully compared with experimental measurements to test the expected relationship between $\tau$ from simulations and tensile strength found experimentally.

The trends observed for the $\tau$ are more difficult to understand than those for $\eta$. There is evidently no clear-cut correlation of $\tau$ with the formation of bridging chains or with the attractive particle-particle interactions. In the absence of such a correlation, we suggest that the nanoparticles effect the mechanical properties by modifying the ratio of the bulk and shear moduli in the low temperature nanocomposite state (i.e. the nanocomposite Poisson ratio). Recent work has noted that molecularly thin sheets have a negative Poisson ratio when they are flexible enough to crumple by thermal fluctuations. Such additives could reduce the Poison ratio of the material as a whole and provide a potential rationale for interpreting the increases in both the strength and toughness observed in our simulation of sheets dispersed in a polymer matrix, as well as in experiments on clay nanocomposites. 

\section{Acknowledgments}
The authors thank V.~Ganesan, S.~Kumar, and K.~Schweizer for helpful 
discussions. We thank the NSF for support under grant number 
DMR-0427239.

\newpage
\begin{table}
\caption{The table details all system variants simulated; listing chainlength $N$, number of chains $N_{\rm C}$, loading fraction $\phi$, number of force sites per nanoparticle $n$, and number of nanoparticles $n_{\rm N}$. }
\begin{tabular}{c|c|c|c|c|c}
System Type~ & $N$ & $N_{\rm C}$ & $\phi$ & $n$ & $n_{\rm N}$ \\
\colrule
Pure~ & ~10~ & ~100~ & 0 & ~N/A~ & ~N/A \\
& 20 & 100 & 0 & N/A & N/A \\
& 40 & 100 & 0 & N/A & N/A \\
\colrule
Icosahedra~ & 10 & 400 & 0.0494 & 13 & 16 \\
& 20 & 200 & 0.0494 & 13 & 16 \\
& 40 & 100 & 0.0494 & 13 & 16 \\
\colrule
Rods~ & 10 & 944 & 0.0505 & 10 & 50 \\
& 20 & 472 & 0.0505 & 10 & 50 \\
& 40 & 236 & 0.0505 & 10 & 50\\
\colrule
sheets~ & 10 & 944 & 0.0505 & 100 & 5 \\
& 20 & 472 & 0.0505 & 100 & 5 \\
& 40 & 236 & 0.0505 & 100 & 5 \\
\end{tabular}
\label{Table1}
\end{table}

\newpage

\end{document}